\documentclass[fleqn]{cas-dc}

\usepackage[numbers]{natbib}

\begin{document}
\let\WriteBookmarks\relax
\let\printorcid\relax
\def\floatpagepagefraction{1}
\def\textpagefraction{.001}

\shorttitle{Mill Valley Evacuation Study}

\shortauthors{D. Pierce and Y. Chen}

\title [mode = title]{Mill Valley Evacuation Study}                      

%
\author[1]{\color{black}Damien Pierce}[type=editor]
\cormark[1]


\affiliation[1]{organization={Google, LLC},
    addressline={1600 Amphitheatre Pkwy},
    city={Mountain View},
    postcode={94043}, 
    state={CA},
    country={United States of America}}

\author[1]{\color{black}Yi-fan Chen}[type=editor]
\cormark[2]


\cortext[cor1]{Principal corresponding author; {\ttfamily dmpierce@google.com}}
\cortext[cor2]{Corresponding author; {\ttfamily yifanchen@google.com}}

\begin{abstract}
Traffic evacuation planning can be essential in saving lives in case of natural disasters such as hurricanes, floods and wildfires. We build on a case study of traffic evacuation planning for the city of Mill Valley, CA. We run a microscopic traffic simulator to examine various evacuation scenarios. We modify some crucial aspects of a previous study to make the simulation more pertinent. For a citywide evacuation, we quantify the importance of decreasing the number of vehicles per household. We find a set of changes that can significantly reduce the evacuation time by routing more traffic to the least-used highway on-ramps. We show results when evacuating various areas of the city one at a time.
\end{abstract}


\maketitle

\section{Introduction}
Weather-related natural disasters are becoming more frequent and over the last decade have affected billions of people ((\citet{UNISDR:2015}). Some communities prepare for natural disasters by developing a traffic evacuation plan. Such a plan can play a crucial role in saving lives.

Given the lack of real-world data, traffic simulations are one of the few viable approaches available to investigate the various aspects of a traffic evacuation plan. We use SUMO (\citet{SUMO:2022}), an open source agent-based traffic simulator, to evaluate the evacuation plan of Mill Valley, California. We first consider the constraints involved and the results obtained in case of a citywide evacuation. We provide a simple way to understand the simulation results in terms of bottlenecks. This leads to a suggested change to the evacuation plan which routes more traffic to the least-used highway on-ramps. This significantly decreases the evacuation time. Next, we show results per area, first during a citywide evacuation, and then considering one area evacuating at a time. We end with some conclusions.
 
\begin{figure}[!t]
  \centering
  \includegraphics[width=\linewidth]{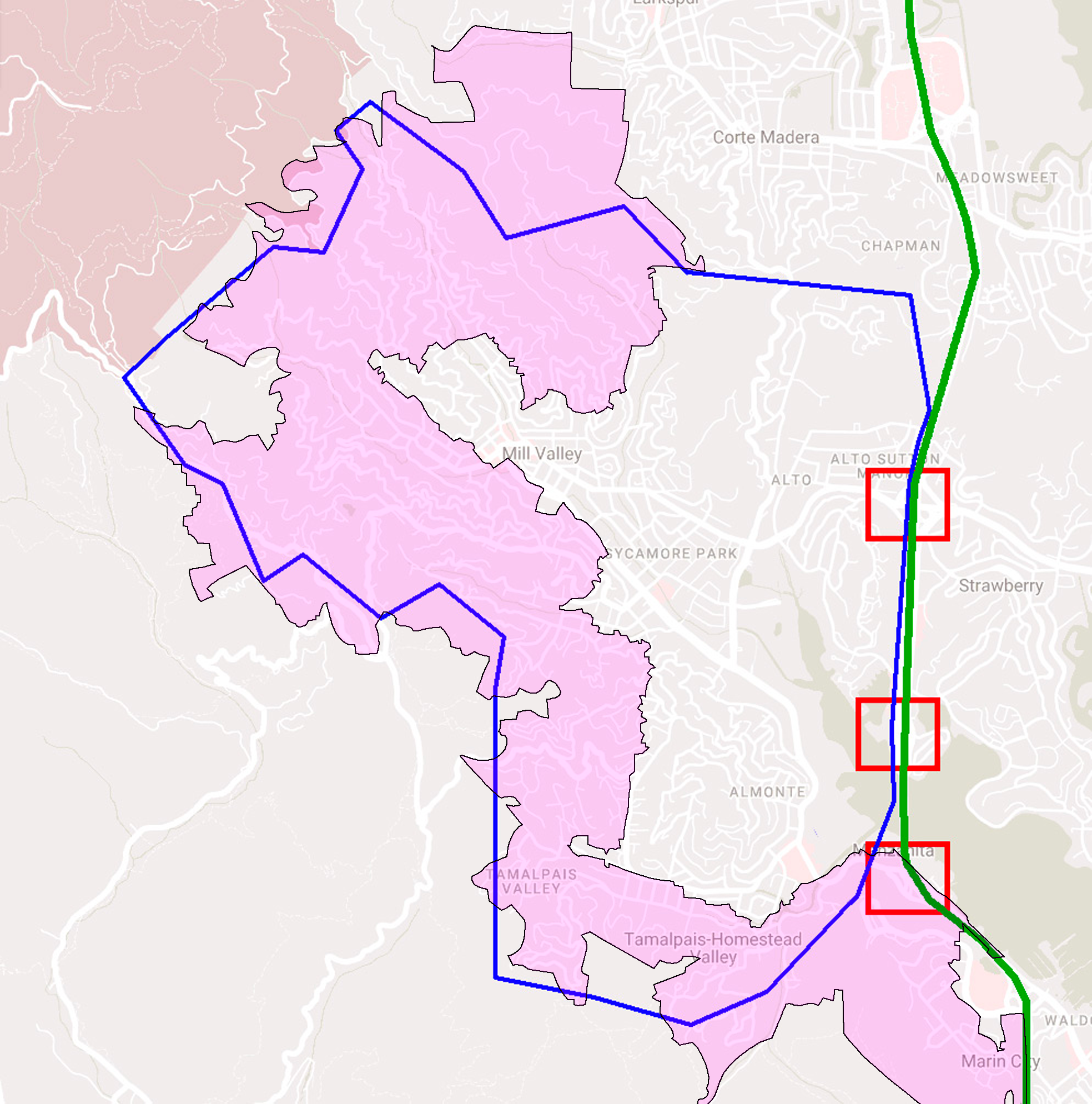}
  \caption{Residences inside the blue outline are included in the simulation (this includes incorporated and unincorporated Mill Valley). Highway 101 is shown in green. The red outlines show the three sets of Highway 101 on-ramps. The purple area has the highest fire hazard rating.}
  \label{fig:mill-valley-region}
\end{figure}
 
Mill Valley is in northern California and is comprised of several narrow valleys and the surrounding hillsides. The hills, both where the houses are situated as well as in the surrounding forests, are dense with trees. Mill Valley's northern, eastern and southern hillsides have the highest fire hazard severity level, according to CAL FIRE, California's fire-fighting agency (\citet{CALFIRE:2022}) (see the light purple area in Fig. \ref{fig:mill-valley-region}). The trees are specially susceptible to fire due to long periods of drought in recent decades. Near the end of 2022, the entire county was designated as being in severe drought according to NOAA NIDIS (\citet{NIDIS:2022}). (The 2022-2023 winter saw record rainfall which has alleviated the drought situation.)

Because of Mill Valley's susceptibility to wildfire, city authorities developed a preliminary evacuation plan which includes changes to the road network designed to help speed up the evacuation. City authorities are responsible for alerting residents and managing the traffic situation in case an evacuation is invoked. Evacuation involves thousands of vehicles, many of which need to navigate narrow windy roads down to a small number of arterials which lead to Highway 101. Highway 101 runs north-south on the eastern side of the city.

\subsection{Mill Valley evacuation overview}

In the following we consider the incorporated and unincorporated parts of Mill Valley west of Highway 101. Mill Valley includes some areas east of 101 but those areas are much closer to the highway and are not as susceptible to fire. The area we consider is shown outlined in blue in Fig. \ref{fig:mill-valley-region}. It has about 11,400 households and each household has on average 2 vehicles. We consider the case where all vehicles start at residences, as would happen in the middle of the night. We assume every household will evacuate in one or more vehicles. We consider scenarios where the average number of evacuating vehicles per household is between 1 and 2. The main questions concerning a citywide evacuation (within the context of the simulation we are doing) are: How quickly can somewhere between 11 and 23 thousand vehicles distributed all over the city get onto Highway 101? What can be done to make the evacuation more efficient?

Fig. \ref{fig:mill-valley-region} shows the three sets of Highway 101 on-ramps (northern, middle, southern) outlined in red. The northern on-ramps have two outgoing lanes approaching them. The middle and southern on-ramps each have one outgoing lane approaching them. Hence, every vehicle must travel on one of four lanes just before entering Highway 101. The most efficient evacuation will be one in which all four lanes are steadily in use during the entire evacuation period. The northern on-ramps have E Blithedale Ave feeding them, and there is a large section of the hills in the northern and the western parts of the city which will route to E Blithedale Ave. Similarly, the southern on-ramps have a large population in the south and southwest that will naturally route to them via the sequential arterial Miller Ave / Almonte Blvd / CA-1. The middle 101 on-ramps do not have a large population naturally routing to them. So unless something is done, the middle 101 on-ramps will run out of traffic well before the evacuation is complete. One goal of this research is to develop a strategy to make better use of the middle Highway 101 on-ramps.

\section{Running SUMO}
We ran version 1.12.0 of the traffic simulator SUMO (\citet{SUMO:2022}) to obtain detailed traffic estimates. SUMO has two basic inputs: a road network and a routes file (with origin/destination pairs, origin times, and route information). Auxiliary files can be included which define things like destination regions, detectors, and map-embedded rerouters.

\subsection{SUMO inputs}
\subsubsection{Network}\label{sec:map}
We start with the same OpenStreetMap-sourced map (\citet{OSM:2022}) of the Mill Valley area used in \citet{CHEN:2020}. We consider this network and two variations. The first has traffic lights disabled. The second is a network modified to include Mill Valley's preliminary evacuation plan, as described below.
\begin{itemize}
\item Traffic lights are disabled.
\item Contraflow is implemented on E Blithedale Ave in the section starting at Lomita and ending near Meadow Dr, so it has two outgoing lanes and no incoming lanes. Incoming emergency vehicles have to take Ashford Ave to Lomita Dr.
\item A small section of Hamilton Dr between just south of the police department and Eucalyptus Knoll St is changed from one-way to two-way (the second lane is not normally used for vehicular traffic). This opens a more direct route from E Blithedale Ave to the middle 101 on-ramps via Roque Moraes and Hamilton.
\item A second outgoing lane is added on Miller Ave from downtown to just past Willow St.
\end{itemize}

Mill Valley city officials have developed these changes to expedite an evacuation. Later in this paper we consider altering this plan by adding one more section of contraflow on E Blithedale Ave between Camino Alto and Roque Moraes (which makes that section have three outgoing lanes and one incoming lane) and restricting traffic behavior in a few places.

\subsubsection{Routes}

The routes file includes the origin and destination for each vehicle, as well as the departing time and route information. One of the refinements we made relative to the previous Mill Valley evacuation study (\citet{CHEN:2020}) is in the number and placement of origins. In the previous study origins were evenly spaced on the residential streets according to an average density. In our case we converted all the house and apartment addresses to geo-coordinates and used public information to take into account the number of households at each address (for example, a single-family home vs. an apartment building). We found 11,372 households in Mill Valley west of Highway 101. This includes households in the incorporated and unincorporated parts of Mill Valley. The previous study only considered about half this number of households, corresponding to the population of the incorporated part of Mill Valley. We checked that our placement of origins at the actual coordinates of residences does not make an appreciable difference compared to the even distribution of origins as used in \citet{CHEN:2020}. However, it is crucially important that we use about twice as many households as that study.

The destinations in the routes file are all set to the same traffic assignment zone (TAZ). The TAZ is comprised of the northernmost and southernmost segments of the section of Highway 101 included in the network. When a vehicle reaches either end of Highway 101, its trip is complete and it is removed from the simulation.

Various studies have found empirically that notification and departure time cumulative distributions in emergency evacuations follow S-shaped curves (\citet{JHA:2004}, \citet{SORENSON:2000}, \citet{WOLSHON:2001}, \citet{OZBAY:2012}, \citet{LI:2013}, \citet{LORETO:2019}, \citet{WOO:2017}, \citet{NOAA:2002}, \citet{PBSJ:2020}). The exact shape of the curve (including, importantly, a measure of the width of the non-cumulative distribution) cannot be known in advance. We use a gamma distribution with a mean of 3 hours and a standard deviation $\sigma = 15$ minutes. With that, the middle 80\% of vehicles leave within about $38$ minutes. In the area-based evacuations, we also consider for comparison an extreme scenario in which all vehicles leave much closer to the same time by using $\sigma = 3$ minutes. In that case 80\% of vehicles leave within about $8$ minutes. In our results and discussion we ignore the initial part of the gamma distribution that has no or almost no traffic.

Since we set the destination in the routes file to a TAZ, when SUMO adds a vehicle to the simulation it assigns an efficient route per the current traffic conditions. Additionally, we enable automatic routing for 75\% of the vehicles, so that most vehicles update their routes once per minute. Using these settings does not imply that 75\% of vehicles are using GPS. Automatic routing helps the vehicles behave more realistically in a general sense. As an example, without automatic routing SUMO can route large amounts of traffic from one arterial to another, even though this eventually causes that route to become congested, while the first arterial is clear all the way to Highway 101. A real driver trying to get to 101 without using GPS would not leave the first arterial in that case.

\subsubsection{Auxiliary inputs and parameters}

We include a few auxiliary SUMO files in our simulation. One of the files defines the destination TAZ (the north and south 101 destination points). Another file defines detectors, which provide information about when vehicles reach certain points of interest (such as the four lanes approaching Highway 101 on-ramps). A file defines map-embedded rerouters. Another auxiliary file specifies polygons, which are used to measure the time it takes for vehicles to reach defined areas. There are many model parameters that affect the simulation. One main one is the step size, which we set to 1 second.

\subsection{SUMO outputs}

Being that the simulation is agent-based, the results include detailed information about each vehicle at each step of the simulation. 

\section{Citywide evacuation results}

We initially consider three networks mentioned above: the original network from OpenStreetMap which we call the baseline network; the same with traffic lights disabled; and the same modified to correspond to the city's preliminary evacuation plan (detailed in Sec. \ref{sec:map}). Later in the paper we will show some results for 1, 1.5, and 2 vehicles per household, but we start by showing results for perhaps the most likely case: 1.5 vehicles per household (on average).

\begin{figure}[htpb]
  \centering
  \includegraphics[width=\linewidth]{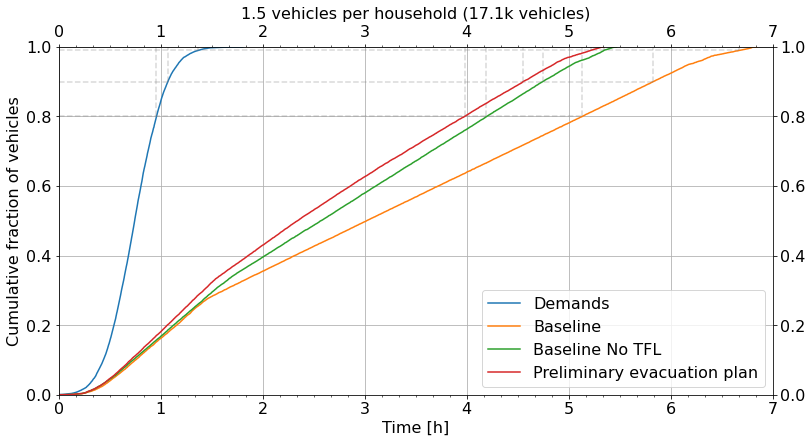}
  \caption{1.5 vehicles per household demand/evacuation curves. The curve labeled ``Baseline'' has traffic lights enabled. ``Baseline No TFL'' is the same with traffic lights disabled. ``Preliminary evacuation plan'' has traffic lights disabled and a few other changes (see Sec. \ref{sec:map}).}
  \label{fig:evac-1.5-cphh}
\end{figure}

Figure \ref{fig:evac-1.5-cphh} shows the demand and evacuation curves for a citywide evacuation simulation with a temporal demand distribution standard deviation $\sigma = 15$ minutes and 17 thousand vehicles. The dashed lines indicate the 80\%, 90\% and 99\% fraction lines and the corresponding times.

The far left blue line in the plot is the cumulative demand curve. This is the requested demand, which is close to but in general differs from the actual departure times. The differences are due to cases where a road becomes so congested that SUMO is not able to add a vehicle at the time and location specified in the routes file. In such cases SUMO will try to add the vehicle every time step until it succeeds. The real-world analogy for this is, some evacuees may find it difficult to exit their driveways because the road is so congested. The other lines are the evacuation curves for the baseline and preliminary evacuation plan cases.

The most striking feature of the plot is that disabling traffic lights significantly decreases evacuation time. The total evacuation time of close to 7 hours (Baseline) drops to a little above 5.5 hours (Baseline No TFL). On the other hand, implementing the city's preliminary evacuation plan changes (see Sec. \ref{sec:map}; the plan includes disabling traffic lights and making rule changes on several road segments) causes only a small decrease in evacuation time relative to the ``Baseline No TFL" case. The total evacuation time decreases from 5:25 to 5:19, a mere 6 minutes difference. It is about 12 minutes difference at the 80\% and 90\% points.

We just referred to some total evacuation times. The large uncertainty in the demand curve transfers directly to the total evacuation time. The time difference between the demand curve and an evacuation curve is less dependent on specifics of the demand curve. In general, when discussing traffic simulation results, time differences have less uncertainty. The differences in hours between the evacuation curve and the demand curve at the 80\%, 90\% and 99\% points are listed in Table \ref{tab:evac-times} for the case of 1.5 vehicles per household. The table also shows the amount the time difference is decreased by, when comparing the current row to the previous row. The evacuation time difference decreases by about 25\% when traffic lights are disabled, and decreases by a further 3 to 7 percent when implementing the other preliminary evacuation plan changes.

\begin{table}[!htbp]
\resizebox{\linewidth}{!}{
\begin{tabular}{@{}lrrrrrr@{}}
\toprule
\multirow{2}{*}{Network} & \multicolumn{2}{c}{80\%} & 
                           \multicolumn{2}{c}{90\%} & 
                           \multicolumn{2}{c}{99\%} \\
& Time (h) & \% diff & Time (h) & \% diff & Time (h) & \% diff \\
\midrule
Baseline & 4.2 & -- & 4.8 & -- & 5.3 & -- \\
Baseline No TFL & 3.2 & 23\% & 3.7 & 23\% & 4.0 & 25\% \\
Preliminary evacuation plan & 3.0 & 6.7\% & 3.5 & 5.2\% & 3.9 & 3.3\% \\
\bottomrule
\end{tabular}}
\caption{Evacuation times in hours for 1.5 vehicles per household. The table shows the difference between the evacuation curve and the demand curve at the 80\%, 90\%, and 99\% evacuation points. The amount the time decreases is also shown in percent, relative to the previous line.}
\label{tab:evac-times}
\end{table}

\subsection{Rate per lane}

Figure \ref{fig:evac-1.5-cphh} shows that there is not much benefit in implementing the preliminary evacuation plan changes beyond disabling traffic lights. Disabling traffic lights is sufficient to create many time periods where one or more of the lanes leading to 101 are saturated. During those periods the lane or lanes are operating near maximum capacity. Improving upstream flows that lead to such a lane cannot decrease the evacuation time much if the lane is already operating near its maximum capacity. This is why the preliminary evacuation plan's road rules changes result in only a modest improvement over the baseline no-traffic-light case.

\begin{figure}[htpb]
  \centering
  \includegraphics[width=\linewidth]{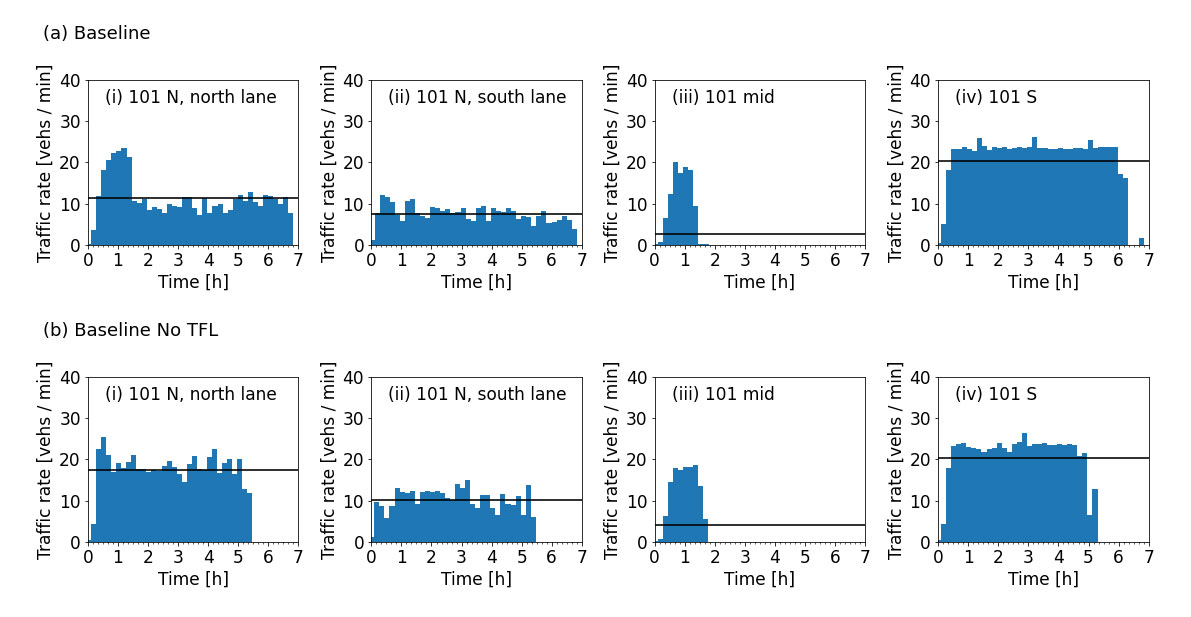}
  \caption{Histograms showing the traffic rate (in vehicles per minute) vs. time for each lane leading to Highway 101 on-ramps with 1.5 vehicles per household. The horizontal lines show the average rate. (a) The Baseline case. (b) The Baseline No TFL case.}
  \label{fig:rates-bl-tfl-no-tfl}
\end{figure}

As mentioned in the introduction, there are three sets of Highway 101 on-ramps. From west of 101, the northern on-ramps have two lanes leading to them and the middle and southern on-ramps have one lane each. Within the context of our simulation, all the traffic from every part of the city has to travel on one of these four lanes to reach Highway 101. In SUMO the rate of traffic of each lane can be measured by adding detectors. Fig. \ref{fig:rates-bl-tfl-no-tfl}(a) shows histograms of the the rate of each of the four lanes in vehicles per minute in the ``Baseline'' case and Fig. \ref{fig:rates-bl-tfl-no-tfl}(b) shows the histograms in the ``Baseline No TFL'' case. In both cases there are on average 1.5 vehicles per household.

The sum of the average rate per lane is a measure of the effectiveness of the evacuation, and is equivalent to the total evacuation time. The rate sum (sum of all four lanes) is 42 vehicles per minute in the Baseline case and 52 vehicles per minute in the Baseline No TFL case. For 17.1k vehicles, these give evacuation times of 6:49 and 5:28, which are within a few of minutes of the results in Fig. \ref{fig:evac-1.5-cphh}.

We were able to check a rate achieved in our simulation against a measured rate. There is a traffic counter on the single-lane-in-each-direction section of E. Blithedale Ave, about 350 yards east of Roque Moraes Dr. This data shows that a typical rate during the peak rush hour is 23 vehicles per minute. In our baseline simulation with traffic lights we measured a peak rate of 22 vehicles per minute at the same spot. This gives us some confidence that the simulation is realistically modelling traffic flow.

In the most efficient evacuation, each lane would have a fairly constant rate over the entire evacuation period, and the average rate of each lane would be close to the maximum rate possible for that lane. Neither of these is the case. Fig. \ref{fig:rates-bl-tfl-no-tfl} shows that the rate on some lanes is far from constant. This is especially true for the middle 101 on-ramps which receive no traffic after the second hour. The average rate per lane varies significantly, being (11, 7.5, 2.5, 20) vehicles per minute in the Baseline case, and (17, 10, 4.0, 20) vehicles per minute in the preliminary evacuation plan case, for the lanes leading to 101-N (N and S), 101-mid, and 101-S on-ramps, respectively.

\subsection{Improving the citywide evacuation}

The rate per lane information can inform strategies for decreasing the overall evacuation time by focusing on the lane or lanes that are underutilized. The lane leading to the southern 101 on-ramps does very well in the preliminary evacuation plan simulation, averaging 20 vehicles per minute. The E Blithedale lanes leading to the northern on-ramps do not do as well. But the big underperforming lane is the one leading to the middle 101 on-ramps which achieves less than 4 vehicles/minute on average over the course of the evacuation. Two factors lead to the poor performance. The first is that the number of residences in the vicinity of those on-ramps (which naturally route to the middle on-ramps) is small. The second reason is that the large volume of traffic on E Blithedale which could route to the middle on-ramps routes instead to the northern on-ramps, as the distance to the northern on-ramps is much shorter.

These observations motivate a set of changes we call the final evacuation plan. We worked with city officials to arrive at changes which are feasible to implement given the available city resources. The changes are as follows:
\begin{itemize}
\item The southernmost lane on the normally westbound section of E Blithedale east of Camino Alto is converted to contraflow (about a 1000 foot section).
\item Lane changing is disabled on the two-lane portion of E Blithedale from Camino Alto to the freeway on-ramps.
\item Northbound traffic on Camino Alto reaching E Blithedale is forced to turn right onto the rightmost lane of E Blithedale and forced to turn right again a block later onto Roque Moraes.
\item Turns off of Roque Moraes are blocked, so that the southbound traffic routes via Hamilton to the middle on-ramps.
\item All traffic on the southern lane of E Blithedale takes the first highway on-ramp. All traffic at the southern on-ramps takes the northern on-ramp to avoid a hard right turn.
\end{itemize}

\begin{figure}[htpb]
  \centering
  \includegraphics[width=\linewidth]{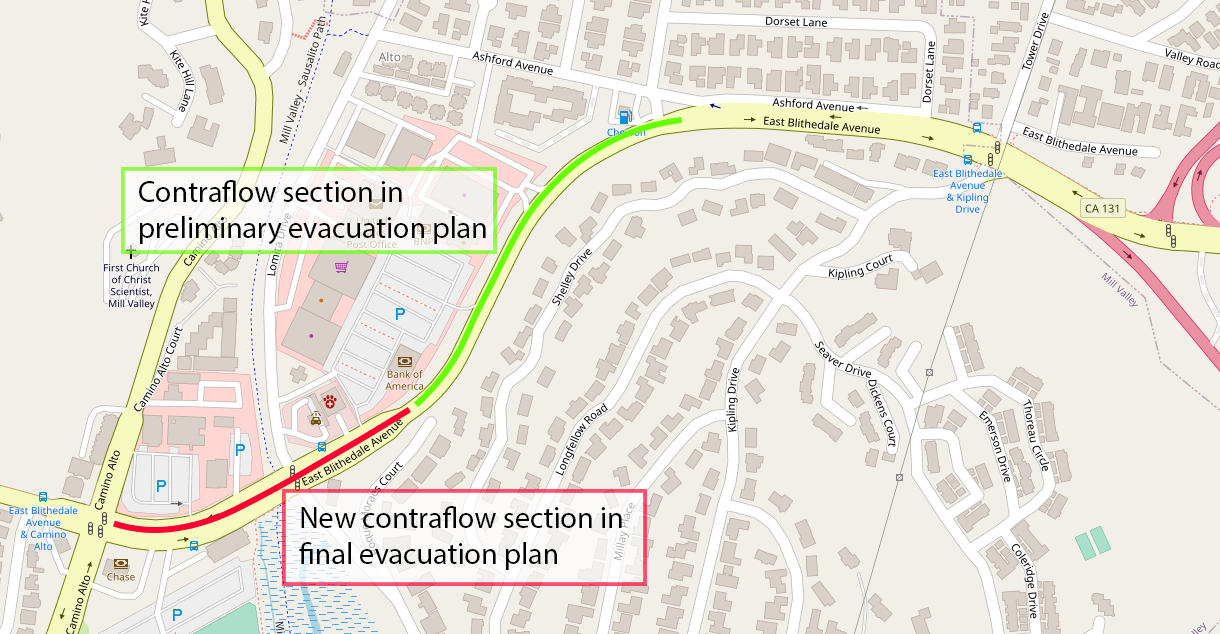}
  \caption{Contraflow sections of E Blithedale Ave. The contraflow section in the preliminary evacuation plan is shown in green. The extra section of E Blithedale converted to contraflow in the final evacuation plan is shown in red.}
  \label{fig:final-contraflow}
\end{figure}

\begin{figure}[htpb]
  \centering
  \includegraphics[width=\linewidth]{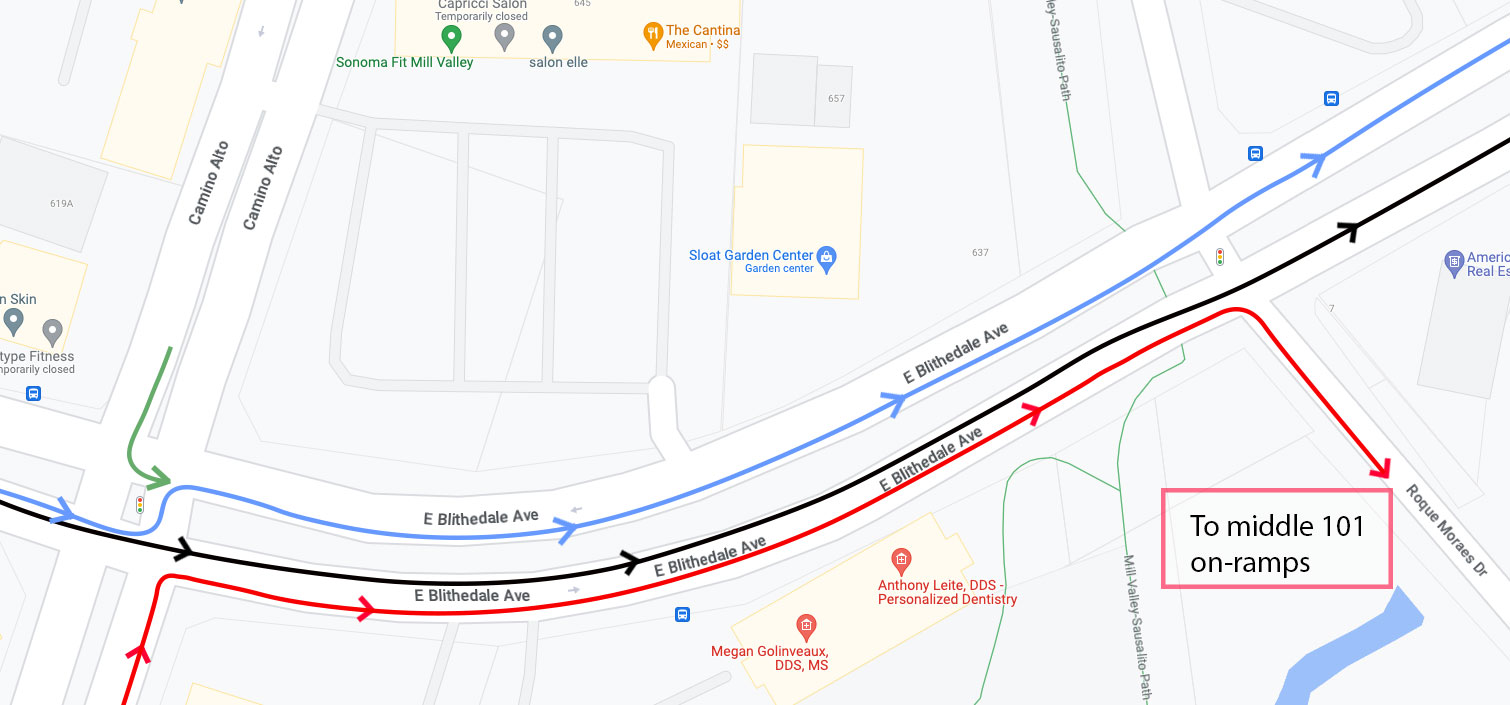}
  \caption{Some routes near the intersection of E Blithedale and Camino Alto in the final evacuation plan. The red line shows traffic heading north on Camino Alto is routed to the middle 101 on-ramps via Roque Moraes. The blue line shows the leftmost lane eastbound on E Blithedale is routed along the new contraflow section of E Blithedale east of Camino Alto. Southbound traffic on Camino Alto that reaches E Blithedale (green arrow) also routes along the new contraflow section.}
  \label{fig:final-intersection}
\end{figure}

The first item is visualized in Fig. \ref{fig:final-contraflow} which shows the part of E Blithedale Ave that is converted to contraflow in the preliminary evacuation plan in green, and the extra part in the final plan in red. Fig. \ref{fig:final-intersection} shows how the left lane on E Blithedale west of the intersection gets routed to the new contraflow section (blue line), as does traffic heading south on Camino Alto (green arrow). The third item is also depicted in Fig. \ref{fig:final-intersection}, which shows how the traffic heading north on Camino Alto towards E Blithedale is routed to the middle on-ramps. The last item speeds up highway ingress by having the traffic take the on-ramps which avoid turns or sharp curves.

Preventing lane changing on E Blithedale is important in order that simulated traffic behaves realistically along the two-lane contraflow section, especially towards the end near the Highway 101 on-ramps. The SUMO route assignments are such that traffic in one lane too often routes to the other lane. Sometimes when this happens in both lanes simultaneously it can cause temporary deadlocks. Hence, the through rate along E Blithedale is markedly increased by preventing lane changes in this section. However, it seems clear that real-life traffic would not behave this way. The main thing we are achieving by preventing lane changing along this segment is getting the traffic to behave realistically. That said, preventing lane changing could provide some benefit, and our results without lane changing could be considered an upper bound on the through rates of the two lanes.

\begin{figure}[htpb]
  \centering
  \includegraphics[width=\linewidth]{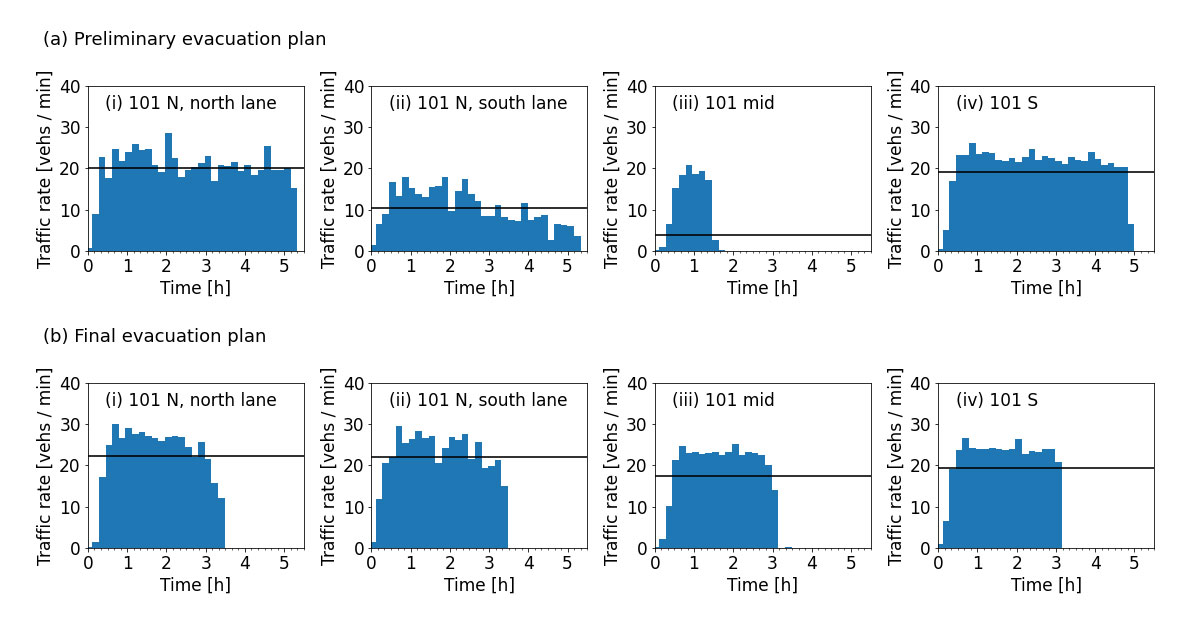}
  \caption{Histograms showing the traffic rate (in vehicles per minute) vs. time for each lane leading to Highway 101 on-ramps with 1.5 vehicles per household. The horizontal lines show the average rate. (a) The preliminary evacuation plan case. (b) The final evacuation plan case.}
  \label{fig:rates-evac-mod-evac}
\end{figure}

These changes are designed to help the evacuation in two ways. First, they create a flow of traffic during most of the simulation to the underutilized middle on-ramps. Second, they create two lanes from the main intersection at E Blithedale and Camino Alto all the way to the northern 101 on-ramps. The two lanes are made independent by preventing lane changes. This further improves each lane's through rate.

Fig. \ref{fig:rates-evac-mod-evac} shows the rate-per-lane comparison of the preliminary and final evacuation plans. The rate on each of the four lanes leading to 101 on-ramps is shown for 1.5 vehicles per household. This plot shows that the final plan is successful in routing significantly more traffic to the middle on-ramps, increasing the average rate from 3.7 to 17.5 vehicles per minute. It also improves the combined rates of the two lanes on E Blithedale from 30 to 43 vehicles per minute.

\begin{figure}[htpb]
  \centering
  \includegraphics[width=\linewidth]{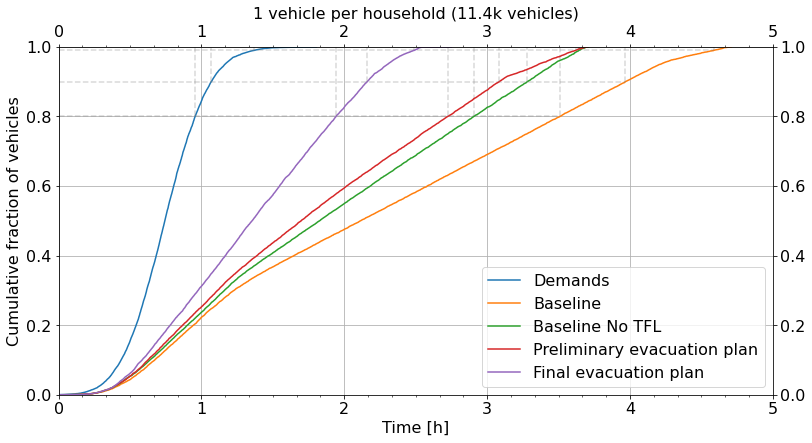}
  \caption{One vehicle per household demand/evacuation curves, as in Fig. \ref{fig:evac-1.5-cphh}, with an additional line showing the result for the final evacuation plan.}
  \label{fig:evac-1-cphh}
\end{figure}

\begin{figure}[htpb]
  \centering
  \includegraphics[width=\linewidth]{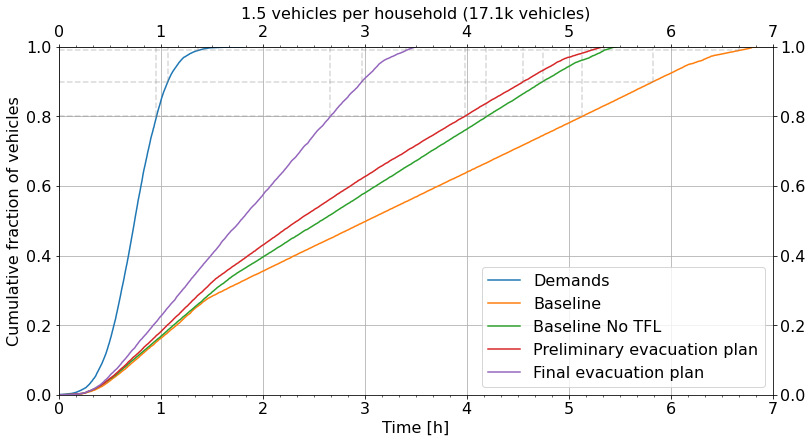}
  \caption{Same as in Fig. \ref{fig:evac-1.5-cphh}, with an additional line showing the result for the final evacuation plan.}
  \label{fig:evac-1.5-cphh-4-maps}
\end{figure}

\begin{figure}[htpb]
  \centering
  \includegraphics[width=\linewidth]{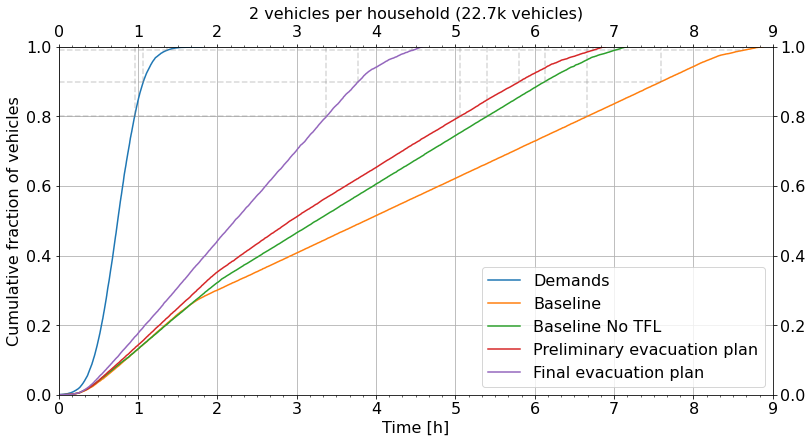}
  \caption{Two vehicles per household demand/evacuation curves, as in Fig. \ref{fig:evac-1.5-cphh}, with an additional line showing the result for the final evacuation plan.}
  \label{fig:evac-2-cphh}
\end{figure}

The evacuation curves are shown in Figs. \ref{fig:evac-1-cphh}, \ref{fig:evac-1.5-cphh-4-maps} and \ref{fig:evac-2-cphh} for one and two vehicles per household, respectively. The final evacuation plan curves show a drastic improvement over the preliminary evacuation plan. The evacuation times are summarized in Table \ref{tab:evac-times-for-varying-vphh}. The differences in hours between the demand and evacuation curves at 90\% are shown for the four networks, for various vehicles per household. The percentage decrease from the previous line is also shown. The final plan lowers the evacuation time difference by over 40\% compared to the preliminary evacuation plan.  

\begin{table}[t]
\resizebox{\linewidth}{!}{%
\begin{tabular}{@{}lrrrrrr@{}}
\toprule
\multirow{2}{*}{Network} & \multicolumn{2}{c}{1 veh/household} & 
                           \multicolumn{2}{c}{1.5 veh/household} & 
                           \multicolumn{2}{c}{2 veh/household} \\
& Time (h) & \% diff & Time (h) & \% diff & Time (h) & \% diff \\
\midrule
Baseline & 2.9 & -- & 4.8 & -- & 6.5 & -- \\
Baseline No TFL & 2.2 & 24\% & 3.7 & 23\% & 5.1 & 23\% \\
Preliminary plan & 2.0 & 8.9\% & 3.5 & 5\% & 4.7 & 6.3\% \\
Final plan & 1.1 & 46\% & 1.9 & 45\% & 2.7 & 43\% \\
\bottomrule
\end{tabular}}
\caption{Time in hours between the demand and evacuation curves at the 90\% point for 1, 1.5 and 2 vehicles per household, and the percentage decrease compared to the previous line.}
\label{tab:evac-times-for-varying-vphh}
\end{table}

\subsection{Per-area citywide evacuation results}

City officials involved in evacuation planning are interested in simulation evacuation results broken down by area.
\begin{figure}[t]
  \centering
  \includegraphics[width=.9\linewidth]{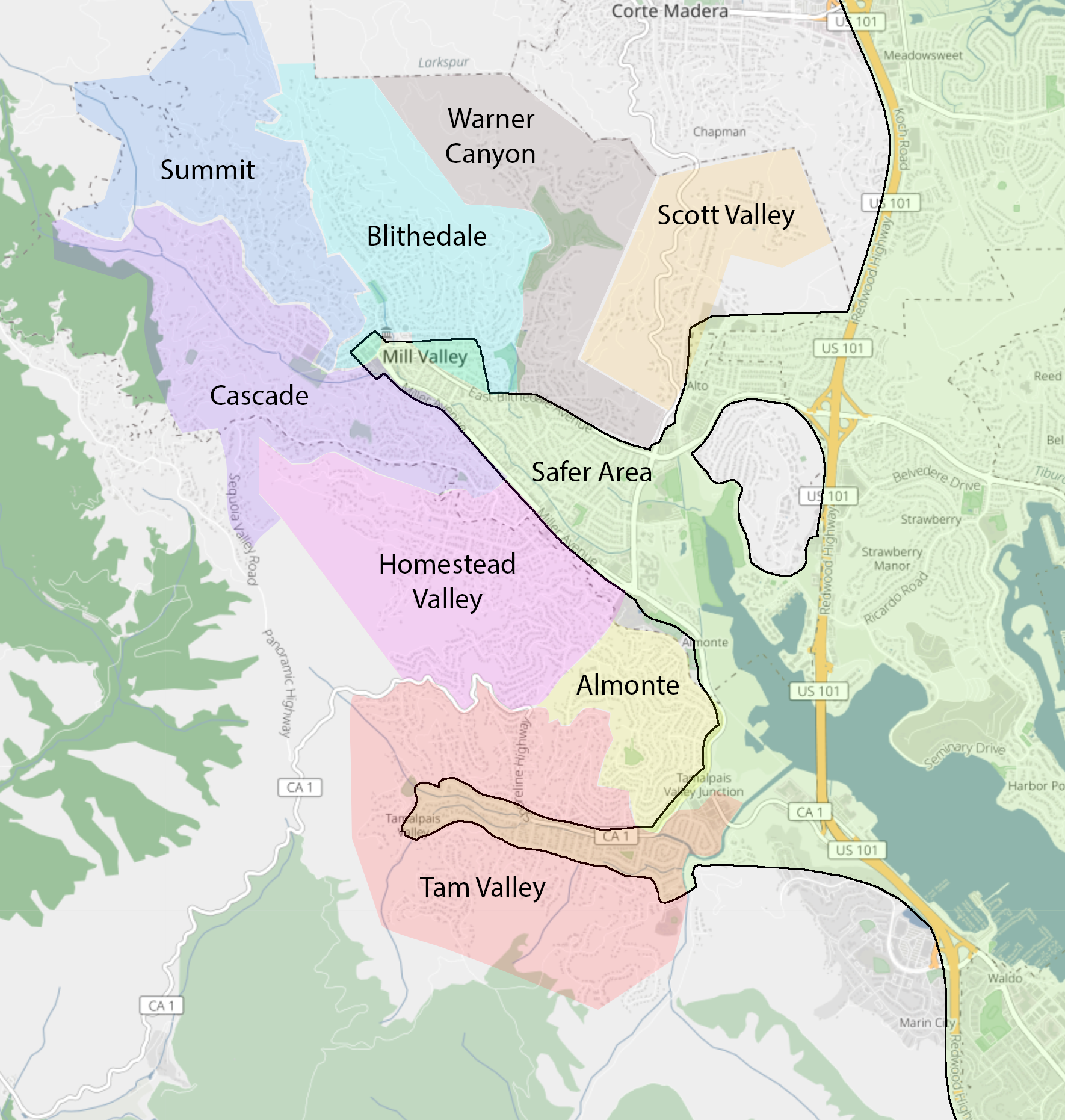}
  \caption{Eight Mill Valley areas. The safer zone is outlined in black and shaded light green. It partially overlaps some named areas.}
  \label{fig:mill-valley-areas}
\end{figure}
This can inform them of the relative vulnerability of each area during an evacuation. A Mill Valley fire official divided up the hilly parts of the city into 8 areas as shown in Fig. \ref{fig:mill-valley-areas}. Any vehicle not originating in one of these 8 areas is in the default area ``Other''. The areas are listed in in Table \ref{tab:areas}, along with the number of households per area. A relatively safer zone was also defined, being the lowest lying, flattest part of the city. This is shown in the green-shaded area with the black outline in Fig. \ref{fig:mill-valley-areas}. The parts of Mill Valley outside of the green safer zone have been identified as a wildland-urban interface (WUI) (\citet{MCFD:2022}). According to the National Park Service, WUI's have an increased likelihood of being impacted by fire because they contain large amounts of plant landscaping, fuel sources and structures that could sustain a fire (\citet{NPS:2022}).

\begin{table}[t]
\begin{tabular}{@{}lrrrrrr@{}}
\toprule
Area & \# of households \\
\midrule
Scott Valley & 269 \\
Summit & 474  \\
Warner Canyon & 482 \\
Almonte & 909  \\
Blithedale & 1188  \\
Cascade & 1272  \\
Homestead Valley & 1325  \\
Tam Valley & 1961 \\
Other & 3492 \\
\bottomrule
\end{tabular}
\caption{The number of households in nine Mill Valley areas.}
\label{tab:areas}
\end{table}

\begin{figure}[t]
  \centering
  \includegraphics[width=0.9\linewidth]{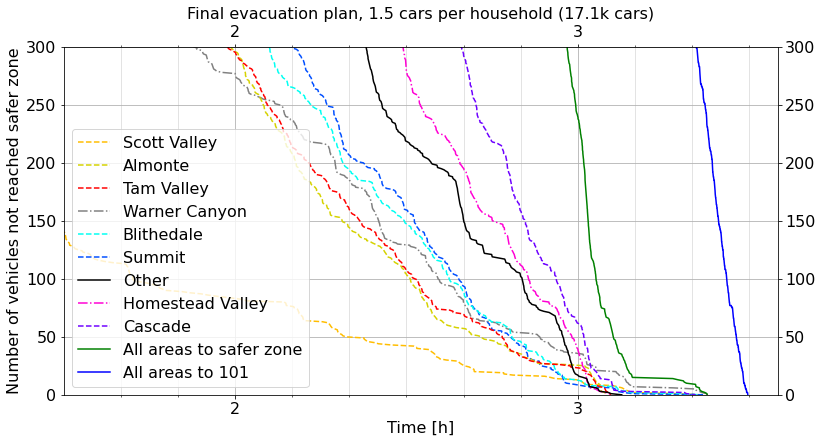}
  \caption{The per area number of vehicles that have not reached the safer zone vs. time for the final evacuation plan with 1.5 vehicles per household. The total for all vehicles not reaching the safer zone is shown in solid green. The total for all areas not reaching 101 is shown in solid blue. The legend is sorted by the curves at the mid-point of the y-axis (150 vehicles).}
  \label{fig:vehicles-per-area-not-reached-safer-zone}
\end{figure}

\begin{figure}[htpb]
  \centering
  \includegraphics[width=0.9\linewidth]{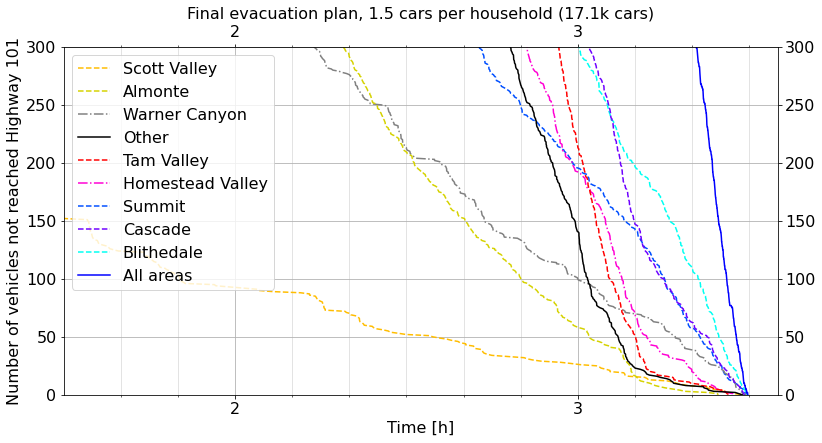}
  \caption{As Fig. \ref{fig:vehicles-per-area-not-reached-safer-zone}, but showing the number of vehicles that have not reached Highway 101.}
  \label{fig:vehicles-per-area-not-reached-highway-101}
\end{figure}

We ran the citywide evacuation simulation keeping track each vehicle according to the area from which it originated. Figs. \ref{fig:vehicles-per-area-not-reached-safer-zone} and \ref{fig:vehicles-per-area-not-reached-highway-101} show the number of vehicles per area that have not reached the safer zone or Highway 101, respectively, vs. the evacuation time, for the final evacuation plan and 1.5 vehicles per household. In Fig. \ref{fig:vehicles-per-area-not-reached-safer-zone} the solid green line shows the total number of vehicles that have not reached the safer zone. In both figures the solid blue line shows the total number of vehicles that have not reached Highway 101. Due to inherent uncertainties in the simulations, the absolute numbers of vehicles in such per-area results should only be used to generally compare areas.

A major factor affecting how quickly an area evacuates is the number of vehicles starting in that area. But other factors such as the average distance traveled and details of the road network also play a role. Scott Valley has the smallest number of vehicles, and its curves are farthest left in both plots. But Summit is the second least populous area, and its curve is mainly sixth from left in the safer zone plot and between third and seventh from left in the Highway 101 case, not second from the left as might be expected. Starting at the other end, Tam Valley has the largest number of vehicles (other than ``Other''), and its curve is mostly seventh from the right in the ``safer zone'' case and fifth from right in the Highway 101 case, rather than second from right as would be expected based only on the number of vehicles. In terms of reaching the safer zone, the slowest areas are Cascade and Homestead Valley. Tables \ref{tab:number-of-vehicles-to-safer-zone} and \ref{tab:percentage-of-vehicles-to-safer-zone} list either the number of vehicles that \emph{have not} reached or the percentage of vehicles that \emph{have} reached the safer zone, per area, for various times.

\begin{table}[!htbp]
\resizebox{.9\linewidth}{!}{%
\begin{tabular}{@{}lrrrrrr@{}}\toprule
\multicolumn{1}{c}{Area} & 1:00 & 1:30 & 2:00 & 2:30 & 3:00 \\
\midrule
Scott Valley & 241 & 141 & 84 & 43 & 13 \\
Summit & 558 & 495 & 371 & 160 & 9 \\
Warner Canyon & 501 & 398 & 278 & 130 & 36 \\
Almonte & 759 & 503 & 297 & 105 & 26 \\
Blithedale & 969 & 584 & 363 & 148 & 13 \\
Cascade & 1453 & 1216 & 848 & 412 & 62 \\
Homestead Valley & 1336 & 1095 & 621 & 279 & 40 \\
Tam Valley & 1167 & 550 & 296 & 107 & 24 \\
Other & 1827 & 641 & 443 & 219 & 17 \\
\midrule
Total & 8811 & 5623 & 3601 & 1603 & 240 \\
\bottomrule
\end{tabular}}
\caption{The number of vehicles in a citywide evacuation that have not reached the safer zone per area at various times, with 1.5 vehicles per household.}
\label{tab:number-of-vehicles-to-safer-zone}
\end{table}

\begin{table}[!htbp]
\resizebox{.9\linewidth}{!}{%
\begin{tabular}{@{}lrrrrrr@{}}\toprule
\multicolumn{1}{c}{Area} & 1:00 & 1:30 & 2:00 & 2:30 & 3:00 \\
\midrule
Scott Valley & 40\% & 65\% & 79\% & 89\% & 97\% \\
Summit & 24\% & 33\% & 50\% & 78\% & 99\% \\
Warner Canyon & 31\% & 45\% & 61\% & 82\% & 95\% \\
Almonte & 45\% & 64\% & 79\% & 92\% & 98\% \\
Blithedale & 46\% & 67\% & 80\% & 92\% & 99\% \\
Cascade & 24\% & 36\% & 56\% & 78\% & 97\% \\
Homestead Valley & 33\% & 45\% & 69\% & 86\% & 98\% \\
Tam Valley & 61\% & 81\% & 90\% & 96\% & 99\% \\
Other & 65\% & 88\% & 92\% & 96\% & 100\% \\
\bottomrule
\end{tabular}}
\caption{The percentage of vehicles in a citywide evacuation that have reached the safer zone per area at various times, with 1.5 vehicles per household.}
\label{tab:percentage-of-vehicles-to-safer-zone}
\end{table}

\section{Area-based evacuations}

Evacuation times will vary in a first approximation proportionally with the number of vehicles. With the final evacuation plan and one vehicle per household, the time difference between the demand curve and the evacuation curve at the 90\% point is 1.1 hours. For two vehicles per household, it is 2.7 hours, or 2.5 times larger, compared to the naive factor of 2. The same ratio for the baseline network is 2.2. The city has already implemented a public awareness campaign to encourage the citizens to evacuate in one vehicle. The actual average number of vehicles per household will not be in the city's control and will not be known in advance. Given a possibly many hours long citywide evacuation, we are motivated to consider area-based evacuations.

If a forest fire forces an evacuation, it is likely that the fire will approach the city from one direction. Of the 8 vulnerable areas defined above, city officials may be able to identify one or two areas that are in the most danger and evacuate them first. The evacuation could then proceed area by area until every area in danger is evacuated.

In an area-by-area evacuation, it can be useful for evacuation officials to have an estimate for how long it will take each area to evacuate, so they know the timing to use when notifying each area in turn. To this end, we run individual simulations for each area. We do not consider background traffic, shadow evacuations, or the effect of one area on another. Also, we are in this case interested in the time it takes vehicles to reach the closest arterial. Once most or all the vehicles from an area have reached arterials, the evacuation of the next area could begin.

Table \ref{tab:individual-areas-data-sigma-t-15} has results for the case of 1.5 vehicles per household and a temporal demand distribution having a standard deviation of $\sigma = 15$ minutes for the final evacuation plan. The table lists the average distance traveled and average time taken to reach the closest arterial, along with the standard deviations. The last two columns show the number of minutes it takes for 90\% or 95\% of vehicles to reach an arterial. These columns show the total time in minutes, and in parentheses the number of minutes beyond the demand distribution time, which is the extra time due to traveling. The total time can include delays which happen when a road is so congested the simulator cannot insert a vehicle at the given time and place.

\begin{table}
\resizebox{\linewidth}{!}{%
\begin{tabular}{@{}lllll@{}}
\toprule
Area & \makecell{Distance \\ \footnotesize{(km)}} & 
\makecell{Time \\ \footnotesize{(mins)}} & \makecell{90\% time \\ \footnotesize{(mins)}} &
\makecell{95\% time \\ \footnotesize{(mins)}} \\
\midrule
Scott Valley & 1.6 \textpm\, 0.6 & 3 \textpm\, 1 & 65 (3) & 72 (3) \\
Summit & 2.1 \textpm\, 1.0 & 4 \textpm\, 2 & 65 (3) & 71 (2) \\
Warner Canyon & 0.9 \textpm\, 0.8 & 1 \textpm\, 1 & 68 (2) & 74 (2) \\
Almonte & 0.9 \textpm\, 0.6 & 3 \textpm\, 4 & 63 (2) & 67 (1) \\
Blithedale & 1.2 \textpm\, 1.1 & 7 \textpm\, 8 & 73 (10) & 77 (9) \\
Cascade & 1.8 \textpm\, 1.3 & 5 \textpm\, 4 & 71 (6) & 75 (4) \\
Homestead Valley & 1.2 \textpm\, 0.8 & 3 \textpm\, 3 & 70 (2) & 75 (2) \\
Tam Valley & 0.9 \textpm\, 0.9 & 5 \textpm\, 7 & 79 (9) & 86 (10) \\
\bottomrule
\end{tabular}}
\caption{Evacuation data for individual area evacuations, for the eight Mill Valley areas with 1.5 vehicles per household and $\sigma = 15$ minutes. These results are for the final evacuation plan, but results for other plans are similar. The 90\% and 95\% columns show the time for 90\% or 95\% of the vehicles to reach an arterial, and, in parentheses, the difference between the total time and the demand distribution time. The time in parentheses is a measure of travel time.}
\label{tab:individual-areas-data-sigma-t-15}
\end{table}

\begin{table}
\resizebox{\linewidth}{!}{%
\begin{tabular}{@{}lllll@{}}
\toprule
Area & \makecell{Distance \\ \footnotesize{(km)}} & 
\makecell{Time \\ \footnotesize{(mins)}} & \makecell{90\% time \\ \footnotesize{(mins)}} &
\makecell{95\% time \\ \footnotesize{(mins)}} \\
\midrule
Scott Valley & 1.8 \textpm\, 0.7 & 7 \textpm\, 5 & 25 (12) & 26 (12) \\
Summit & 2.8 \textpm\, 1.4 & 13 \textpm\, 7 & 32 (19) & 34 (19) \\
Warner Canyon & 1.4 \textpm\, 1.1 & 7 \textpm\, 6 & 28 (13) & 29 (13) \\
Almonte & 1.4 \textpm\, 1.0 & 10 \textpm\, 9 & 31 (17) & 36 (21) \\
Blithedale & 2.6 \textpm\, 2.4 & 21 \textpm\, 15 & 54 (40) & 57 (41) \\
Cascade & 2.7 \textpm\, 2.1 & 13 \textpm\, 11 & 40 (26) & 42 (26) \\
Homestead Valley & 1.5 \textpm\, 1.2 & 9 \textpm\, 8 & 33 (19) & 37 (21) \\
Tam Valley & 1.1 \textpm\, 1.2 & 15 \textpm\, 16 & 56 (38) & 61 (33) \\
\bottomrule
\end{tabular}}
\caption{Same as Table \ref{tab:individual-areas-data-sigma-t-15}, but with $\sigma = 3$ minutes.}
\label{tab:individual-areas-data-sigma-t-3}
\end{table}

So far we assumed a temporal demand distribution with a width $\sigma$ of $15$ minutes. This leads to demand times at 90\% of about 65 minutes. If the demand is more concentrated, congestion effects can play a more significant role in travel times. As an extreme case we consider $\sigma = 3$ minutes. For this distribution 90\% of vehicles depart within about 16 minutes. In this case we find the increased congestion leads to travel times that are markedly larger than in the case $\sigma = 15$ minutes. However, this is more than compensated for by the decrease in spread of departure times, so that 95\% of vehicles reach an arterial in less time overall. The shortest 95\% time in Table \ref{tab:individual-areas-data-sigma-t-15}, 67 minutes, is longer than the longest 95\% time in Table \ref{tab:individual-areas-data-sigma-t-3}, 61 minutes. Hence, if everyone leaves from a given area within a short time window, even though they will experience more congestion and longer trip times, the area will still be evacuated more quickly than when the evacuation time is dominated by the departure distribution.

\section{Conclusion}
We present the results of a case study of citywide and area-based evacuations of Mill Valley, CA, such as might be ordered in the event of a nearby wildfire. We used an agent-based traffic simulator to model the evacuation of between 11 and 23 thousand vehicles from Mill Valley, CA to Highway 101.

In the case of a citywide evacuation, we show results for the baseline network and three other networks. The city of Mill Valley had a preliminary plan to enact a set of changes to facilitate a citywide evacuation. These include disabling traffic lights, and changing road rules in several places. We first consider only disabling traffic lights and find large ($\sim25\%$) reductions in evacuation times. We then consider the remaining set of changes in the city's preliminary plan. These include changing road rules to allow for additional lanes in a couple places and implementing contraflow in one section of an arterial. We do not find much change due to these changes compared to only disabling traffic lights.

In our simulation, all traffic must travel over one of four lanes in order to reach Highway 101. We look at the number of vehicles per minute each lane yields over the course of the simulation. We find one of the lanes does not receive much traffic and, working with the city officials, we develop changes in one block of the road network to send more traffic to that lane. With these road changes, we find over 40\% additional reduction in evacuation times compared to the preliminary evacuation plan. The city has adopted the final evacuation plan.

In the last section of the paper we consider per-area evacuations. We first show citywide evacuation results broken down by area, then consider evacuating one area at a time. We provide statistics for each area including average time and distance traveled to reach an arterial, and the total time for 90\% or 95\% of vehicles to reach an arterial. We compare results for two very different time distributions. We find that in case all vehicles depart in a short time window, evacuees will encounter more congestion leading to longer travel times, but the overall time to reach an arterial is less than a case where the departure distribution dominates the evacuation time.

Results obtained using a traffic simulator are at best indicative. The uncertainties inherent in using a simulator should be kept in mind, especially results involving absolute times or numbers of vehicles. There are a long list of assumptions that go into any simulation. To give one example, we only consider the case that all vehicles start at residences, as would happen in the middle of the night. We also do not have any background traffic in the city or on the highway. We do not consider minor or major road blockages. Additionally, there are unknowns in the inputs (such as the departure time distribution), and in physical and unphysical parameters used to run the simulation. An example of an unphysical parameter is the amount of time a vehicle is stuck before it is teleported to a point further along its route. Such teleportation is needed in order that spurious unrealistic deadlocks do not stop portions of the traffic for unreasonably long periods. Notwithstanding all these points, comparing two simulations that use the same parameters should provide relatively more reliable results.

\section*{Acknowledgements}
We thank Mayor John McCauley and City of Mill Valley personnel Tom Welch, Lindsay
Haynes, Danielle Staude, Rick Navarro and Alan Piombo for numerous discussions
and feedback. Map images in Figs. 1, 4, and 10 from \href{openstreetmap.org/copyright}{OpenStreetMap}.

\printcredits

\bibliographystyle{cas-model2-names}

\bibliography{bibliography}

\end{document}